\documentclass{ws-ijmpd}

\begin{document}

\markboth{Yongli Ping, Lixin Xu, Chengwu Zhang, and Hongya Liu}
{Dark Energy in Global Brane Universe}
%
\catchline{}{}{}{}{}
%
\title{Dark Energy in Global Brane Universe}

\author{Yongli Ping\footnote{ylping@student.dlut.edu.cn}, Lixin Xu, Chengwu Zhang and
 Hongya Liu\footnote{Corresponding author: hyliu@dlut.edu.cn.}
}

\address{School of Physics and Optoelectronic Technology, Dalian
University of Technology, Dalian, Liaoning 116024, P.R.China.}

\maketitle

\begin{abstract}
We discuss the exact solutions of brane universes and the results
indicate the Friedmann equations on the branes are modified with a
new density term. Then, we assume the new term as the density of
dark energy. Using Wetterich's parametrization equation of state
(EOS) of dark energy, we obtain the new term varies with the
red-shift $z$. Finally, the evolutions of the mass density parameter
$\Omega_2$, dark energy density parameter $\Omega_x$ and
deceleration parameter $q_2$ are studied.
\end{abstract}


\section{Introduction}
It is proposed that our universe is a 3-brane embedded in a
higher-dimensional space.\cite{Arkani-Hamed}\cdash\cite{Randall1} In
this brane world model, gravity can freely propagate in all
dimensions, while standard matter particles and forces are confined
on the 3-brane. A five-dimensional ($5D$) cosmological model and
derived Friedmann equations on the branes are considered by
Binetruy, Deffayet and Langlois (BDL),\cite{BDL} for a recent
review, it can be seen.\cite{Marttenreview,Philippe Brax} Recent
observations indicate that our universe is accelerating
\cite{Riess,Perlmutter} and dominated by a negative pressure
component dubbed dark energy. Obviously, a natural candidate is the
cosmological constant with equation of state $w_{\Lambda}=-1$. But,
comic observations imply that dark energy may be
dynamic.\cite{dynamic}\cdash\cite{dynamic2} So, a lot of dark energy
models are studied extensively, such as quintessence, phantom,
etc.\cite{dynamic}\cdash\cite{phantom} While, brane-world models of
dark energy are studied\cite{Sahni2} and accelerating universe comes
from gravity leaking to extra dimension.\cite{Deffayet} In this
paper, we study the dark energy and the universe evolution on the
branes which are embedded in a Ricci-flat bulk characterized by a
class of exact solutions. The solutions were firstly presented by
Liu and Mashhoon and restudied latter by Liu and
Wesson.\cite{L-M,Wesson} And they are algebraically rich because
they contain two arbitrary functions of time $t$. Then they are
studied as Ricci-flat universe widely\cite{X-W}\cdash\cite{ZHANG}
and exact global solutions of brane universes.\cite{Liu}

In this paper, we discuss dark energy in global brane universes. The
exact global solutions of brane universes are studied and they show
that that the Friedmann equations on the second brane ($y\neq0$) are
modified with a new density term. Then we assume this term as the
density of dark energy. Since the EOS of dark energy has been
presented and investigated widely,\cite{Corasaniti}\cdash\cite{Ren}
now we use the Wetterich's parametrization EOS of dark
energy\cite{Wetterich} to study the dark energy on the brane
$y\neq0$ (the second brane). This paper is organized as follows: In
Section II, we derive the Friedmann equations with a new term on the
second brane from the exact global solutions and assume this new
term as the density of dark energy. In Section III, the evolutions
of the dimensionless density parameters of matter $\Omega_2$ and
dark energy $\Omega_x$ respectively and deceleration parameter $q_2$
on the branes $y\neq0$ are obtained by using Wetterich's
parametrization EOS of dark energy to study dark energy. Section IV
is a short conclusion.

\section{Friedman equations in brane universes}
The $5D$ cosmological solutions\cite{L-M,Wesson} read
\begin{equation}
dS^{2}=B^{2}dt^{2}-A^{2}\left( \frac{dr^{2}}{1-kr^{2}}+r^{2}d\Omega
^{2}\right) -dy^{2}, \label{line element}
\end{equation}%
where $d\Omega ^{2}=d\theta ^{2}+\sin^2 \theta d\psi ^{2}$ and $k$
is the $3D$ curvature index ($k=\pm 1,0$). For the solutions satisfy
the $5D$ vacuum equation $R_{AB}=0$, they are used as the bulk
solutions of BDL-type brane model. To obtain brane models for using
the $Z_2$ reflection symmetry on $A$ and $B$, they are set
as\cite{Liu}
\begin{eqnarray}
A^{2}&=&\left( \mu ^{2}+k\right) y^{2}-2{\nu}\mid{y}\mid+\frac{\nu
^{2}+K}{\mu ^{2}+k} ,  \nonumber\\
B&=&\frac{1}{\mu }\frac{\partial A}{\partial t}\equiv
\frac{\dot{A}}{\mu },  \label{A}
\end{eqnarray}%
where $\mu =\mu \left( t\right) $ and $\nu =\nu \left( t\right) $
are two arbitrary functions, and $K$ is a constant.

Then the corresponding $5D$ bulk Einstein equations are taken as
\begin{eqnarray}
G_{AB}&=&\kappa^2_{(5)}T_{AB}, \nonumber\\
T_B^A&=&\delta(y)diag(\rho_1,-p_1,-p_1,-p_1,0)  \nonumber \\
&\ &+\delta(y-y_2)diag(\rho_2,-p_2,-p_2,-p_2,0)\label{G}
\end{eqnarray}
where the first brane is at $y=y_1=0$ and the second is at
$y=y_2>0$. In the bulk $T_{AB}=0$ and $G_{AB}=0$, Eq.(\ref{G}) are
satisfied by (\ref{A}). On the branes Liu had solved Eq.(\ref{G}) in
Ref.~\refcite{Liu}. We adopt the result at $y=y_1=0$ and $y=y_2>0$
as follows:
\begin{eqnarray}
\kappa^2_{(5)}{\rho}_1&=&\frac{6\nu}{A_1^2}, \label{rho1}\\
\kappa^2_{(5)}p_1&=&-\frac{2}{\dot{A_1}}\frac{\partial}{\partial{t}}(\frac{\nu}{A_1})-\frac{4\nu}{A^2_1}
\label{p1},
\end{eqnarray}
and
\begin{eqnarray}
\kappa^2_{(5)}{\rho}_2&=&\frac{6}{A_2}(\frac{\mu^2+k}{A_2}y_2-\frac{\nu}{A_2}), \label{rho2}\\
\kappa^2_{(5)}p_2&=&-\frac{2}{\dot{A_2}}\frac{\partial}{\partial{t}}(\frac{\mu^2+k}{A_2}y_2-\frac{\nu}{A_2}) \nonumber\\
&\ &-\frac{4}{A_2}(\frac{\mu^2+k}{A_2}y_2-\frac{\nu}{A_2})
\label{p2}.
\end{eqnarray}
Now, we consider the universe on the second brane, i.e. $y=y_2>0$.
From the $5D$ metric (\ref{line element}), the Hubble and
deceleration parameters on $y=y_2$ brane can be defined as
\begin{eqnarray}
H_2(t,y)&\equiv&\frac{1}{B_2}\frac{\dot{A_2}}{A_2}=\frac{\mu}{A_2},
\label{H}\\
q_2(t,y)&=&-\frac{A_2\dot{\mu}}{\mu\dot{A_2}}.\label{q_2}
\end{eqnarray}
Substituting Eq.(\ref{H}) into Eq.(\ref{rho2}) to eliminate $\mu^2$,
we find that Eq. (\ref{rho2}) can be rewritten into a new form as
\begin{equation}
H^2_2+\frac{k}{A^2_2}=\frac{\kappa^2_{(5)}}{6y_2}({\rho}_2+\frac{6}{\kappa^2_{(5)}}\frac{\nu}{A^2_2}).
\label{friedman e}
\end{equation}
Comparing with the standard Friedman equation i.e.
$H^2+\frac{k}{A^2}=\frac{\kappa^2_{(4)}}{3}{\rho}$, we can find
${\kappa^2_{(4)}}={\kappa^2_{(5)}}/{(2y_2)}$. Since
$\kappa^2_{(5)}=M^{-3}_{(5)}$ and $\kappa^2_{(4)}=M^{-2}_{(4)}$, the
relation of four dimensional Planck mass is expressed with five
dimensional Planck mass as
\begin{equation}
M^2_{(4)}=2y_2M^3_{(5)}.
\end{equation}
Therefore, the four dimensional Planck mass is relevant to five
dimensional Planck mass and the position of brane. This Friedmann
equation is different from BDL's because it is derived from the
exact solution of $5D$ vacuum equation $R_{AB}=0$ on $y\neq0$ brane.

We assume the term $\frac{6}{\kappa^2_{(5)}}\frac{\nu}{A^2_2}$ as
the density of dark energy. That is,
\begin{equation}
\rho_x=\frac{6}{\kappa^2_{(5)}}\frac{\nu}{A^2_2}. \label{rhox}
\end{equation}
From the Eq.(\ref{p2}), it is obtained
\begin{equation}
\frac{2\mu\dot{\mu}}{A_2\dot{A_2}}+\frac{\mu^2+k}{A^2_2}
=-\frac{\kappa^2_{(5)}}{2y_2}(p_2+p_x),\label{f2}
\end{equation}
where
$p_x=-\frac{2}{\kappa^2_{(5)}}(\frac{\dot{\nu}}{A_2\dot{A}_2}+\frac{\nu}{A^2_2})$.
This is the pressure of dark energy. Then, from Eq. (\ref{q_2}), Eq.
(\ref{friedman e}) and Eq. (\ref{f2}), for $k=0$, the deceleration
parameters $q_2$ can be obtained
\begin{equation}
q_2=\frac{1}{2}\left[\frac{3(p_2+p_x)}{\rho_2+\rho_x}+1\right].\label{q2}
\end{equation}
Meanwhile, the conservation law $T^B_{A;B}=0$ gives
\begin{equation}
\dot{\rho_2}+3(\rho_2+p_2)\frac{\dot{A_2}}{A_2}=0.
\label{conservation}
\end{equation}

\section{Density parameters and their evolution}

From the definement of $\rho_x$ and $p_x$, for $k=0$, we obtain the
EOS of dark energy
\begin{equation}
w_x=\frac{p_x}{\rho_x}=-\frac{1}{3}(\frac{A_2\dot{\nu}}{\dot{A_2}\nu}+1).
\label{EOS}
\end{equation}
For a given component, which has the equation of state
$p_2=w_2\rho_2$ (with $w_2$ being a constant), we obtain
$\rho_2=\rho_{20}A_2^{-3(1+w_2)}$ from Eq.(\ref{conservation}). Now,
considering the given component as matter, i.e. $w_2=0$, we get
$\rho_2=\rho_{20}A_2^{-3}$. Therefore, the dimensionless density
parameters are obtained
\begin{eqnarray}
{\Omega}_2&=&\frac{\rho_{20}}{\rho_{20}+\frac{6}{\kappa^2_{(5)}}{\nu}A_2}, \label{omega-m}\\
{\Omega}_x&=&1-{\Omega}_2,\label{omega-x}
\end{eqnarray}
here $\rho_{20}$ is the current values of matter.

In terms of Eq.(\ref{A}) $A_2$ is a function of $t$ and $y$.
However, on the given $y=y_2$ brane, $A_2$ becomes $A_2=A_2(t)$.
Furthermore, we use the relation
\begin{equation}
A_2=\frac{A_{20}}{1+z}, \label{A(z)}
\end{equation}
and define $\nu=\nu_0f(z)$, and then we find that Eqs.
(\ref{EOS})-(\ref{omega-x}) can be expressed via redshift $z$ as
\begin{eqnarray}
w_x&=&\frac{(1+z)}{3f(z)}\frac{df(z)}{dz}-\frac{1}{3},\label{w-x} \\
\Omega_2&=&\frac{\Omega_{20}(1+z)}{\Omega_{20}(1+z)+(1-\Omega_{20})f(z)}, \label{omz}\\
\Omega_x&=&1-\Omega_2 .\label{oxz}
\end{eqnarray}
From Eqs. (\ref{w-x})-(\ref{oxz}), the evolution of cosmic
components will be determined if the function $f(z)$ is given.

\begin{figure}
\begin{center}
\includegraphics[angle=0,width=3.0in]{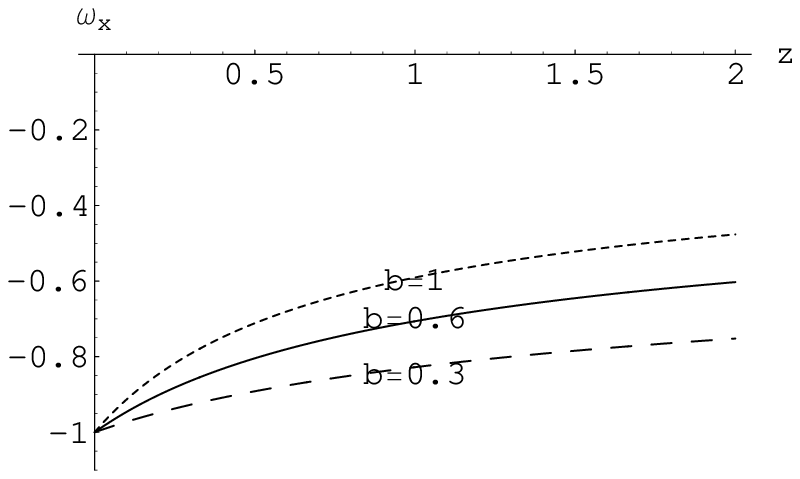}
\end{center}
\caption{$w_x$ of the dark energy as a function of the redshift $z$
with its current value $w_0=-1$ and the bending parameter
$b=1,0.6,0.3$ respectively.}\label{w}
\end{figure}

Now we utilize the form of EOS of the dark energy given by
Wetterich,\cite{Wetterich} which has been
studied.\cite{Gong}\cdash\cite{Wu} The form is
\begin{equation}
w_x(z,b)=\frac{w_0}{1+b\ln(1+z)},\label{wxb}
\end{equation}
where $w_x(z,b)$ is the EOS parameter with its current value as
$w_0$, and $b$ is a bending parameter describing the deviation of
$w_x$ from $w_0$ as $z$ increases. By substituted Eq.(\ref{wxb})
into Eq. (\ref{w-x}), the function $f(z)$ is obtained as follows:
\begin{equation}
f(z)=(1+z)[1+b\ln(1+z)]^{3w_0/b}. \label{fz}
\end{equation}
Substituting Eq. (\ref{fz}) into Eq. (\ref{w-x}), we can obtain
(\ref{wxb}). The only difference is $b\neq0$ in this condition. So,
there must be deviation from $w_0$ as redshift $z$. We plot the
evolution of the EOS of dark energy with $w_0=-1$ in Fig.\ref{w},
where $b=0.3,0.6,1$, respectively. In this figure, we can see $w$
varies with redshift $z$ and $b$. With the increase of $b$, the
decline of $w$ becomes fast.

\begin{figure}
\begin{center}
\includegraphics[angle=0,width=3.0in]{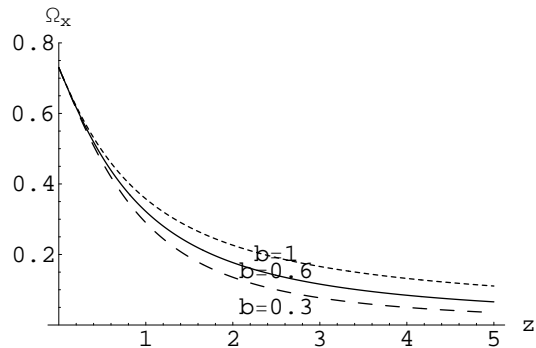}
\end{center}
\caption{Evolution of dark energy $\Omega_x$  vs. redshift $z$ with
$w_0=-1$, $\Omega_{20}=0.27$, $\Omega_{x0}=0.73$ and the bending
parameter $b=1,0.6,0.3$ respectively.}\label{Omega_x}
\end{figure}

Substituting the function Eq. (\ref{fz}) into Eq. (\ref{omz}) and
Eq. (\ref{oxz}), we obtain that
\begin{eqnarray}
\Omega_2&=&\frac{\Omega_{20}}{\Omega_{20}+(1-\Omega_{20})(1+b\ln(1+z))^{3w_0/b}}, \label{omfz}\\
\Omega_x&=&1-\Omega_2 .\label{oxfz}
\end{eqnarray}
where, $\Omega_{20}$ is the current value at $z=0$. From Eq.
(\ref{omfz}) and Eq. (\ref{oxfz}), we can obtain the evolution of
density parameter $\Omega_2$ and $\Omega_x$. Now $\Omega_x$ is
described in Fig.\ref{Omega_x} and it is shown that $\Omega_x$
increases with decrease of redshift $z$ and the larger $b$ is, the
faster $\Omega_x$ increases.

\begin{figure}
\begin{center}
\includegraphics[angle=0,width=3.0in]{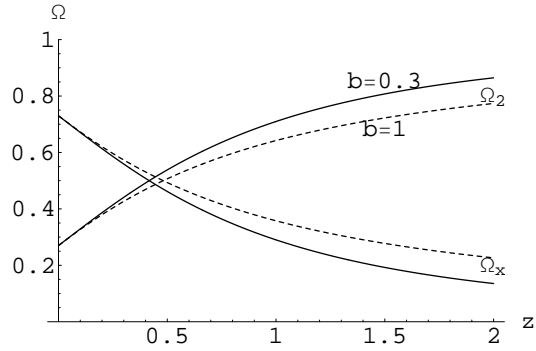}
\end{center}
\caption{Evolution of $\Omega_x$ and $\Omega_2$ vs. redshift $z$
with $w_0=-1$, $\Omega_{20}=0.27$, $\Omega_{x0}=0.73$ and the
bending parameter $b=1,0.3$.}\label{Omega_xm}
\end{figure}

We plot the evolutions of $\Omega_2$ and $\Omega_x$ with redshift
$z$ in Fig. \ref{Omega_xm} where $b=0.3,1$. It is found that two
lines intersect at one point. This point is equilibrium of
$\Omega_2$ and $\Omega_x$ and it varies with $b$, i.e. the
corresponding redshift at the point increases with the growing of
$b$.

\begin{figure}
\begin{center}
\includegraphics[angle=0,width=3.0in]{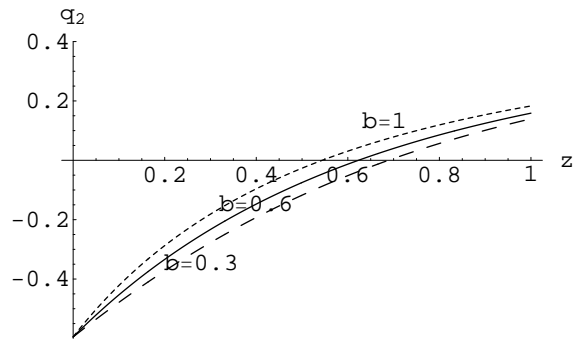}
\end{center}
\caption{Evolution of the deceleration parameter $q_2$ vs. redshift
$z$ with $w_0=-1$, $\Omega_{20}=0.27$, and the bending parameter
$b=1,0.6,0.3$ respectively.}\label{q}
\end{figure}

Also, we obtain the deceleration parameter $q_2$ form Eq. (\ref{q2})
\begin{equation}
q_2=\frac{1}{2}\{\frac{3w_0[1+b\ln(1+z)]^{(\frac{3w_0}{b}-1)}}{\frac{\Omega_{20}}
{1-\Omega_{20}}+[1+b\ln(1+z)]^{\frac{3w_0}{b}}}+1\}.
\end{equation}
The evolution of deceleration parameter $q_2$ with redshift $z$ is
plotted in Fig. \ref{q}. The larger $b$ is, the faster the
attenuation of $q$ is. The transition from decelerated expansion to
accelerated expansion can be seen easily and the redshift $z$ of the
point at $q=0$ becomes smaller with the increase of $b$.

\section{Conclusions}
 The exact global solutions of brane universes are studied.
The solutions contain two arbitrary functions $\mu$ and $\nu$.
 In this paper, we study the Friedmann equation modified on the brane, and the term with $\nu$ in the Friedmann
 equation can drive our universe to accelerate. We assume this term
 with $\nu$ as the density of dark energy on $y\neq0$ brane.
 If different $\nu$ is given, different models of dark
 energy can be obtained. We suppose only matter on the brane i.e.
 $p_2=0$. Using Wetterich's parametrization equation of state
of dark energy and the relation $A_2=A_{20}/(1+z)$, we obtain the
relation of $\nu$ with redshift $z$. Thus, if the current values of
the two density parameters $\Omega _{20}$, $\Omega _{x0}$, $w_{0}$
and the bending parameter $b$ are known, the arbitrary $\nu$ could
be determined uniquely, and then $\mu (z)$ could be\ determined too.
In this way, the whole $5D$ solutions could be reconstructed. And,
in principle the $5D$ solutions could provide with us a global brane
cosmological model to simulate our real universe. We have also
studied the evolutions of matter density $\Omega
_{2}$, dark energy density $%
\Omega _{x}$ and deceleration parameter $q$ with redshift $z$ and
different $b$. These cosmic parameters depend on the bending
parameter $b$. Therefore, we expect accurate observational
constraints on current cosmic parameters and bending parameter $b$
in order to determine the evolution of $5D$ global brane universe.

\section*{Acknowledgments}
This work was supported by NSF (10573003), NSF (10647110), NSF
(10703001), NBRP (2003CB716300) of P. R. China and DUT 893321.


\end{document}